\documentclass{he_symp}
\usepackage{psfig}

\begin{document}
\title{Future Pulsar Observations with H.E.S.S.}
\author{A. Konopelko}  
\institute{Max--Planck--Institut f\"ur Kernphysik, D-69029 Heidelberg, Germany}
\maketitle

\begin{abstract}
Since their discovery at radio wavelengths pulsars have been persistent 
targets for widespread multi-wave observations throughout optics, radio,  
X-rays, and high-energy $\gamma$-rays. Observations with the EGRET $\gamma$-ray 
telescope, on board Compton GRO satellite, confirmed the expectation of 
a pulsed high-energy emission up to a few GeV. Presently, at least seven
objects are known as well established high-energy $\gamma$-ray pulsars. 
A few of those emit $\gamma$-rays well above 1~GeV. Forthcoming ground-based 
\v{C}erenkov telescopes will enable observations of $\gamma$-rays well below
$100$~GeV, finally reaching the yet unexplored energy gap at tens of GeV. H.E.S.S. 
({\it High Energy Stereoscopic System}) is one of such instruments which is planned 
to be operational in 2004. Here I summarize the basic scientific motivations, 
the H.E.S.S. sensitivity, and the first targets for future pulsar observations 
at high energies from the ground. 
\end{abstract}

\section{Introduction}
It is generally accepted that pulsars are produced from massive progenitor 
stars with a masses of $M \geq 8~M_{\odot}$ and typical radii  
$R_o \simeq 10^{11}~\rm cm$, collapsing at the end of their evolution to neutron stars 
in an explosion, which ejects the outer envelope forming a supernova remnant. A pulsar 
is a rotating neutron star (Baade \& Zwicky 1934) with a mass of $M \simeq 1.4~M_{\odot}$, 
a radius of $R \simeq 15$~km, and a rotational period typically of order $P\simeq 1$~s. 
Pulsars are slowing down and loosing their rotational energy with the rate  
\begin{equation}
\dot{E}= 4\pi^2 I \dot{P}/P^3
\end{equation}
where $\dot{P}=dP/dt$ is the measured rate of a pulsar slow-down, and $I=10^{45} \rm g~cm^2$
is a generic moment of inertia of a pulsar. Assuming that the pulsar releases its 
mechanical energy in form of magnetic dipole radiation (Ostriker \& Gunn 1969) one 
can estimate the pulsar surface magnetic field 
\begin{equation}
B_s \cong 3.2\cdot 10^{19} (P \dot{P})^{1/2}, \label{field}
\end{equation}
which appears to be very strong and for a typical pulsar is of the order of $10^{12}\rm G$. 
Assuming a constant magnetic field throughout the pulsar lifetime, the characteristic age of
a pulsar can be determined from its rotational period and slow-down rate
\begin{equation}
\tau = P/(2\dot{P}).\label{tau}
\end{equation}
The $\gamma$-ray luminosity of a pulsar, $L_\gamma \,\, \rm erg\, s^{-1}$, can be 
roughly estimated using the measured flux of high-energy $\gamma$-rays, 
$F_\gamma \, \rm photon\, cm^{-2} \, s^{-1}$:
\begin{equation} 
L_\gamma(>E_\gamma)  \simeq \omega d^2 F_\gamma(>E_\gamma)E_\gamma 
\end{equation}
where $\omega$ is the solid angle into which the pulsar beams and $d$ is the distance to 
the pulsar. The $\gamma$-ray luminosity is widely supposed to be proportional to the 
total loss of pulsar rotational energy
\begin{equation}
L_\gamma \simeq \epsilon \dot{E}
\end{equation}
where $\epsilon$ is the efficiency of converting the rotational energy of neutron star 
into $\gamma$-rays. On the other side the $\gamma$-ray flux of a particular pulsar can 
be estimated as  
\begin{equation}
F_\gamma \propto \epsilon ( \dot{E}/d^2 ){E_\gamma}^{-1} \,\, \rm ph \, cm^{-2} \, s^{-1}.
\label{epsi}
\end{equation}
The conversion efficiency, $\epsilon$, could be, in zero approximation, assumed as a 
phenomenological function of pulsar age, even though this function is not unique and 
not universal. The observed $\gamma$-ray flux depends at first on the size of the pulsar 
beam. Finally the spin-down powered pulsars can be characterized by a set of five 
parameters $(P, \dot{P}, B_s, \tau, \dot{E}/d^2 )$, which allow to predict approximate 
$\gamma$-ray luminosity of a particular pulsar for a given $\dot{E}/d^2$ and $\epsilon$.
More accurate calculations of $\gamma$-ray fluxes, light-curves and energy spectra can 
be performed using  elaborated models of spinning down pulsars.         

\section{Models of high-energy pulsars}

There are two major classes of models dealing with high-energy $\gamma$-ray 
emission from the pulsars. They differ primarily by the initial assumptions on where 
the particle acceleration and radiation take place, either near the surface of neutron star 
at its {\it polar caps} or in the {\it outer gaps}, which may occur in the outer magnetosphere 
of a pulsar at relatively large distance from the surface of the neutron star (see 
Figure~1).  

\begin{figure}
\centerline{\psfig{file=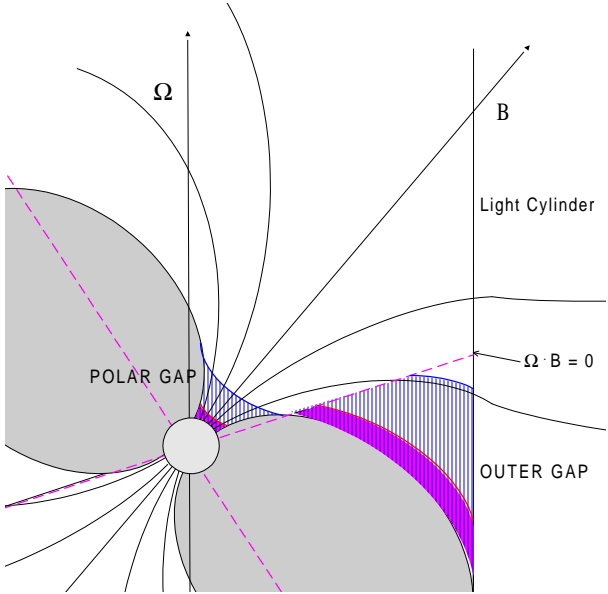,width=8cm,clip=} }
\caption{Schematic view onto the pulsar environment. Courtesy of A.K.~Harding (NASA/GSFC). 
\label{f1}}
\end{figure}
\vspace{2 mm}

\noindent
{\it Polar cap model}. A fast spinning, magnetized neutron star can generate a huge 
electric field, which induces an electrostatic potential  
$U \propto P^{-3/2}(\dot{P})^{1/2} \cong 10^{12}\rm V $ right above the polar caps 
of a pulsar. As suggested by Goldreich \& Julian (1969) the corresponding electrostatic 
force may overcome gravity and pull out the charged particles from the surface 
of the neutron star, forming the current given by $I_{GJ} \propto B_s/P^2$. In 
addition charged particles may escape from the surface of the neutron star because 
its surface temperature $T_s = 10^5 - 10^6 \rm K$ (Becker \& Tr\"{u}mper 1997) may exceed 
the electron thermal emission temperature. The electrons can be accelerated in the regions 
above the poles where the open magnetic field lines\footnote{magnetic field lines, 
which are crossing the light cylinder of a radius $R_{LC} = c/\Omega$, where $\Omega$ is 
an angular velocity of a pulsar.} originate (Sturrock 1971; Ruderman \& Sutherland 1975). Those 
electrons may emit high-energy $\gamma$-rays in three major processes: {\it (i)} synchrotron 
emission induced by electrons spinning around the magnetic field lines; {\it (ii)} 
curvature radiation generated by the electrons moving along magnetic field lines;
{\it (iii)} inverse Compton emission produced by high-energy electrons, which 
up-scatter low energy photons to $\gamma$-ray energies. The $\gamma$-rays interact 
with the pulsar magnetic field ($\gamma \rightarrow e^- e^+$) initiating the pair cascade
(Harding 1981; Daugherty \& Harding 1982). The pair formation fronts occur at 0.5-1 
stellar radii above the surface of the neutron star where the magnetic field is rather 
strong (for a typical pulsar $B\simeq 10^{12}\rm G$). For such magnetic fields only 
the $\gamma$-rays below a certain energy (a few GeV for typical pulsars) may freely 
escape from the pair production region. The super-exponential cut-off in the high-energy 
$\gamma$-ray spectrum is an intrinsic feature the of polar cap model.     

\vspace{2 mm}
\noindent
{\it Outer Gap model}. A vacuum gap may occur in the outer magnetosphere of a pulsar 
(Cheng, Ho \& Ruderman 1986; Romani 1996), which is limited by the null charge surface 
${\bf \Omega \cdot B} \equiv 0$ and the light cylinder. Charged particles can flow out 
through the light cylinder, whereas they could not run through the null charge 
surface. The $\gamma$-rays (curvature or inverse Compton photons) penetrating into the 
vacuum gap can produce a pair ($\gamma\gamma \rightarrow e^- e^+$) by interacting with 
non-thermal X-rays or infrared photons, which fill the gap. The secondary electrons can 
be accelerated in the vacuum gap electric field and emit synchrotron and curvature 
$\gamma$-ray photons.  The $\gamma$-rays of energy at least up to 10~GeV may escape from the 
vacuum gap. This model is suggestive of pulsed TeV $\gamma$-ray emission induced by the 
inverse Compton up-scattering of infrared photons by high-energy electrons (Romani 1996).    
\begin{table}
\caption{High-energy $\gamma$-ray pulsars.}
\label{tb1}
\[
\begin{array}{crcccccc}
\hline
\noalign{\smallskip}
Pulsar  & P (ms) & -log \dot{P} & log{B_s} & log \tau & d~(kpc) & log \dot{E}/d^2 & R^1 \\
\noalign{\smallskip}
\hline
\noalign{\smallskip}
B0531+21   & 33  & 12.4 & 12.6 & 3.1 & 2.0 & -4.9 & 1 \\
B0833-45   & 89  & 12.9 & 12.5 & 4.1 & 0.5 & -5.5 & 2 \\
B1706-44   & 102 & 13.0 & 12.5 & 4.2 & 1.8 & -7.0 & 3 \\
B1509-58^2 & 150 & 11.8 & 13.2 & 3.2 & 4.4 & -7.0 & 4 \\
B1951+32   & 39  & 14.2 & 11.7 & 5.0 & 2.5 & -7.2 & 5 \\
J0633+1746 & 237 & 14.0 & 12.2 & 5.5 & 0.2 & -7.3 & 6 \\
B1055-52   & 197 & 14.2 & 12.0 & 5.7 & 1.5 & -8.9 & 29 \\
\noalign{\smallskip}
\hline
\end{array}
\]
\begin{list}{}{}
\item[$^1$] R is a rank given according to the derived value of $\dot{E}/d^2$. 
\item[$^2$] $PSR~B1509-58$ was seen only with COMPTEL (Kuiper et al. 1999) up 
            to an energy of about 30~MeV, and not with EGRET.
\end{list}
\end{table} 
    
Even though both of the models can fit the variety of light-curves of radio pulsars 
as well as the photon spectra over the optical, radio, X-ray, and high-energy wavelengths,
they inevitably predict very different spectra at energies above 10~GeV. The harder 
spectra of high-energy $\gamma$-rays are more plausible for outer gap model, whereas 
such spectra are almost excluded in the polar cap model (see Harding 2000) 
except for millisecond pulsars. Future pulsar 
observations from the ground with \v{C}erenkov telescopes will allow to measure the 
spectral shape of a pulsed emission in the relevant energy range (above 10~GeV), which 
will be tightly constraining the choice of an appropriate model of high-energy $\gamma$-ray 
emission. 

There are a few other models, which do not fit into the two classes mentioned above, for 
instance a {\it neutron star wind model} suggested recently by Kirk, Skjaeraasen, Gallant (2002). 
The energy of the pulsar magnetic field may be converted into energy of accelerated particles 
within the magnetic field reconnections, which do occur at the null current sheet 
extending rather far from the surface of the neutron star beyond the light cylinder. 
Spectra of $\gamma$-ray emission computed using such a model might be 
challenging future observations at high-energies from the ground.         
\begin{table}
\caption{Spectral parameters$^1$ of high-energy $\gamma$-ray pulsars$^2$. }
\label{tb2}
\[
\begin{array}{lrcrc}
\hline
\noalign{\smallskip}
Object  & K (\times 10^{-8}) & \alpha & E_o & \beta  \\
        & (cm^{-2}s^{-1}GeV^{-1}) &  & (GeV) & \\
\noalign{\smallskip}
\hline
\noalign{\smallskip}
B0531+21   &  24.0 & 2.08 & 30 & 2.0 \\
B0833-45   & 138.0 & 1.62 & 8  & 1.7 \\
B1706-44   &  20.5 & 2.10 & 40 & 2.0 \\
B1951+32   &   3.8 & 1.74 & 40 & 2.0 \\
J0633+1746 &  73.0 & 1.42 & 5  & 2.2 \\
B1055-52   &   4.0 & 1.80 & 20 & 2.0 \\
\noalign{\smallskip}
\hline
\end{array}
\]
\begin{list}{}{}
\item[$^1$] see for details de~Jager et al. (2001)
\item[$^2$] Spectral references from Macomb \& Gehrels (1999) 
\end{list}
\end{table} 
\section{EGRET pulsars}
The discovery of the first radio pulsar $PSR~B1919+21$ (Hewish et al. 1968) has triggered 
further large-scale search for radio pulsars, which brought us a lot more detections. 
These days the catalogs of radio pulsars account for more than 1400 objects. About 
50 of those have been seen in X-rays, but only 7 at high energies ($E_\gamma \geq 30$~MeV)
(Kanbach 2002), recently observed with EGRET\footnote{EGRET is the 
high-energy $\gamma$-ray telescope on board {\it Compton Gamma Ray Observatory (CGRO)}}.
It is remarkable that most of the EGRET pulsars have got a top rank when they are 
listed according to the value of $\dot{E}/d^2$ (see Table~1).   

Most of the high-energy $\gamma$-ray pulsars have been seen at radio wavelengths,
except Geminga ($PSR~J0633+1746$), which appears to be radio-quiet. The light-curves of pulsed 
high-energy $\gamma$-ray emission generally do not resemble the light-curves of radio 
emission, even though they can be rather similar, e.g. for the Crab pulsar ($PSR~B0531+21$). 
Three pulsars, Crab, Vela ($PSR~B0833-45$), and Geminga, have a double-peaked 
light-curve with a substantial contribution of a bridge (inter-peak) emission. The peculiar 
behavior of the light-curves reveals a complicated geometry of the pulsed emission, which 
is evidently not always similar to the light-curves at other wavelengths (for detailed 
discussion see Thompson 2000).

Six $\gamma$-ray pulsars show a clear evidence of the pulsed emission above 5~GeV.
Spectral energy distributions (SED) of $\gamma$-ray emission from the high-energy 
pulsars ($\nu F_{\nu},\,\, \rm Jy\, Hz$) peak in the hard X-ray or $\gamma$-ray range. The SED breaks 
at GeV energies, which is an established  distinctive feature of $\gamma$-ray emission from 
the pulsars, which was predicted by the polar cap model. 

The energy spectra of EGRET pulsars can be well fitted by power laws with spectral indices 
within the range $\alpha$=1.4--2.1 plus a super-exponential cut-off at the energy about 
$E_o = 5-40~\rm GeV$ (de~Jager et al. 2001)
\begin{equation}
dF_\gamma/dE = K \cdot E^{-\alpha}{\rm exp}(-(E/E_o)^{\beta}),\,\, \beta \simeq 2
\label{fit}
\end{equation}    
where $E$ is in units of GeV.
Parameters of the fit for six EGRET pulsars are summarized in Table~2. 

The normalization constant $K$ in Eqn.(\ref{fit}) corresponds to the photon flux 
at energy of 1~GeV, and is an approximate measure of the pulsar luminosity in high-energy 
$\gamma$-rays, $L_\gamma$. One can expect that $L_\gamma$ correlates with the total 
loss of the rotational energy of a pulsar ($\dot{E}/d^2$). In Figure~2 the scatter plot 
of $K$ over $\dot{E}/d^2$ for six GeV EGRET high-energy $\gamma$-ray pulsars is shown. Even 
though the current sample of high-energy $\gamma$-ray is very poor one can see an evident 
trend, which can be fitted as $L_\gamma \propto (\dot{E}/d^2)^{1/3}$.  
\begin{figure}
\centerline{\psfig{file=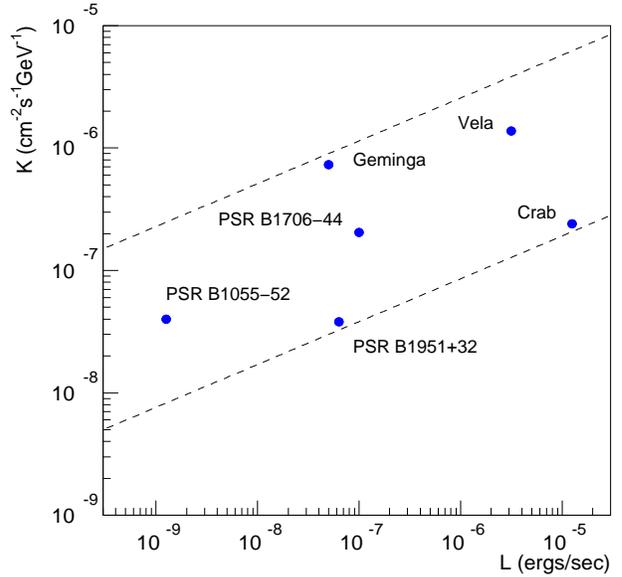,width=8.0cm,clip=} }
\caption{The $\gamma$-ray photon flux at 1~GeV versus rescaled total loss of rotational 
energy $L = \dot{E}/d^2$.  
\label{f1}}
\end{figure}
Pulsar luminosity in high-energy $\gamma$-rays is determined not only by the total loss of 
its rotational power but in addition by the efficiency of energy conversion from slow-down 
into high-energy $\gamma$-rays (see Eqn.(\ref{epsi})). In fact, even a pulsar with a rather low 
value of $\dot{E}/d^2$ and low Rank (see Table~1) could appear to be a powerful $\gamma$-ray 
emitter. For instance $\gamma$-ray pulsar $PSR~B1055-52$ has rather low Rank -- 29. Naturally the 
efficiency $\epsilon = L_\gamma/L \propto K/(\dot{E}/d^2)$ could be a function of pulsar age. 
\begin{figure}
\centerline{\psfig{file=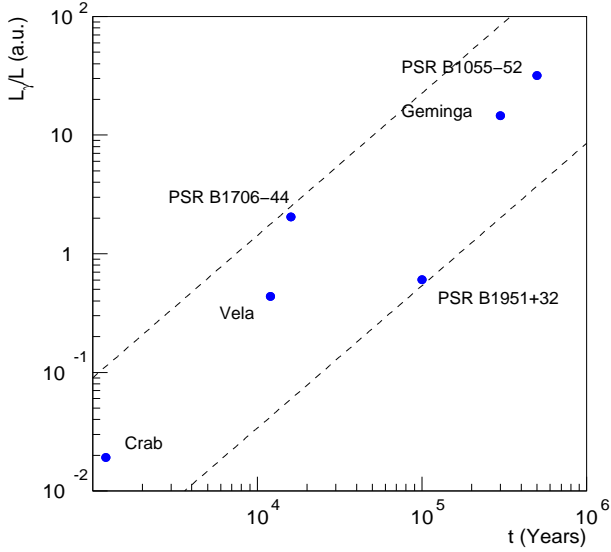,width=8.0cm,clip=} }
\caption{Efficiency of energy conversion of a rotational energy of a pulsar into 
high-energy $\gamma$-ray emission.
\label{f1}}
\end{figure}
The efficiency of energy conversion from rotational slow-down into high-energy $\gamma$-rays 
can be determined at most by the increase of current flow out of the neutron star for elder 
pulsars (Harding 1981), or by the corresponding increase in size of a particle acceleration 
region in the magnetosphere around the star (Ruderman, Cheng 1988). The measure of conversion 
efficiency as a function of pulsar apparent age is shown in Figure~3. The general tendency 
can be approximated by $\epsilon \propto \tau^{1.2}$.

For a better understanding of pulsar workings a substantially more representative sample of 
$\gamma$-ray pulsars is needed. Future observations from space with GLAST\footnote{GLAST: 
The Gamma ray Large Area Space Telescope} and from the ground with future \v{C}erenkov 
instruments may significantly increase the sample of $\gamma$-ray pulsars.

de~Jager et al (2001) have estimated the minimal exposure for a 5$\sigma$ detection of EGRET 
high-energy $\gamma$-ray pulsars. Pulsar $PSR~B1706-44$ appeared to be the best candidate for the 
future observations with H.E.S.S. (for a description of the experiment, see Hofmann 2001). A 10~hours 
exposure with H.E.S.S. will enable to resolve a pulsed high-energy $\gamma$-ray emission 
above 30~GeV from $PSR~B1706-44$. 

\section{Pulsar observations at very high energies}
Based on the outer gap model of $\gamma$-ray emission (see discussion above) Romani (1996) 
predicted a pulsed very high energy (VHE) TeV emission through the inverse Compton (IC) up-scattering 
of synchrotron photons on the primary electrons accelerated in the strong electric field of a pulsar. 
It motivated numerous observations of pulsars with ground-based imaging and non-imaging 
atmospheric \v{C}erenkov telescopes (Ong 1998; Catanese \& Weekes 1999). Note that the 
ground-based instruments are approaching the energy range covered by satellite detectors 
and already nowadays achieve an energy threshold as low as 60~GeV.  Basic results of 
these observations are summarized in Table~2. Except for recent evidence for a detection   
of Geminga with still low level of confidence (Neshpor \& Stepanian 2001) none of the 
observations have resolved a pulsed VHE $\gamma$-ray emission. The outer gap model could easily 
accommodate the non-detection of pulsed VHE $\gamma$-ray emission by assuming higher
fluxes of the microwave seed photons which cause the pair production of TeV $\gamma$-rays 
within the vacuum gap (Hirotani \& Shibata 2001). Predicted fluxes of pulsed IC TeV $\gamma$-rays 
rely heavily on the spectral energy distribution of the microwave seed photons, which is currently
purely known. 
\begin{table}
\caption{Summary of a search for pulsed VHE $\gamma$-ray emission from the ground.}
\label{tb3}
\[
\begin{array}{llllc}
\hline
\noalign{\smallskip}
Pulsar  & Telescope: & E_{th} (GeV) & U.L.^1 & Ref.: \\
\noalign{\smallskip}
\hline
\noalign{\smallskip}
B0531+21  & HEGRA      & 10^3 & \leq 5.6 \times 10^{-13} & \cite{daum} \\ 
          & Whipple    & 250  & \leq 4.8 \times 10^{-12} & \cite{less} \\
          & STACEE     & 190  & \leq 1.2 \times 10^{-11} & \cite{oser} \\
          & CELESTE    &  60  & \leq 7.4 \times 10^{-11}  & \cite{naro} \\
\hline
B0833-45  & CANGAROO   & 2.5\times10^3 & \leq 3.7 \times 10^{-13} & \cite{velc} \\
          & Durham     & 300           & \leq 1.3 \times 10^{-11} & \cite{veld} \\
\hline
B1706-44  & Durham     & 300  & \leq 7.8 \times 10^{-12} & \cite{1706} \\
\hline
B1951+32  & Whipple    & 260  & \leq 6.7 \times 10^{-12} & \cite{srin} \\
\hline
J0633+1746& HEGRA      & 1.5\times10^3 & \leq 1.0 \times 10^{-12} & \cite{daum} \\
          & Whipple    & 500           & \leq 6.5 \times 10^{-12} & \cite{aker} \\ 
          & CrAO^2   & 10^3          & \simeq 2.4 \times 10^{-11}&\cite{nesh}  \\

\hline
B1055-52  & Durham     & 300  & \leq 6.8 \times 10^{-11} & \cite{veld} \\
\noalign{\smallskip}
\hline
\end{array}
\]
\begin{list}{}{}
\item[$^1$] U.L. is an upper limit measured in $\rm photon \, cm^{-2} \, s^{-1}$. 
\item[$^2$] Neshpor \& Stepanian (2001) claim the detection of Geminga pulsar at 
            4.4$\sigma$ confidence level. 
\end{list}
\end{table} 
A detailed search for the IC component of VHE $\gamma$-ray emission from pulsars 
will be performed with the ground-based imaging \v{C}erenkov telescopes of the next 
generation with drastically improved flux sensitivity.   

\section{Plerions}
What sets the puzzle is that the Crab Nebula (Hillas et al. 1998; Aharonian et al. 2000), Vela 
(Yoshikoshi et al. 1997; Chadwick et al. 2000), and $PSR~B1706-44$ (Kifune et al. 1995; 
Chadwick et al. 1998), are well established sources of DC GeV-TeV $\gamma$-ray emission 
in the 60~GeV - 50~TeV energy range. This steady flux emission might be associated possibly 
with the X-ray synchrotron nebula around the pulsar (de~Jager \& Harding. 1992; 
de~Jager~et~al. 1996; Aharonian,
Atoyan, Kifune 1997). However the spectral behavior of both, pulsed and unpulsed, components 
of high-energy $\gamma$-rays within the energy range from 10~GeV up to 100~GeV remains unknown.
This energy range can be effectively studied from the ground with the forthcoming imaging 
atmospheric \v{C}erenkov light telescopes (see for a review Catanese, Weekes 1999).
Spectral measurements over this energy range will severely constrain the physics of  
particle injection and acceleration in a pulsar and within the surrounding pulsar nebula.
Note that H.E.S.S. could detect a Crab like source of DC VHE $\gamma$-rays after a few 
minutes of observation (Konopelko 2000). 
\begin{figure*}[htbp]
\centerline{\psfig{file=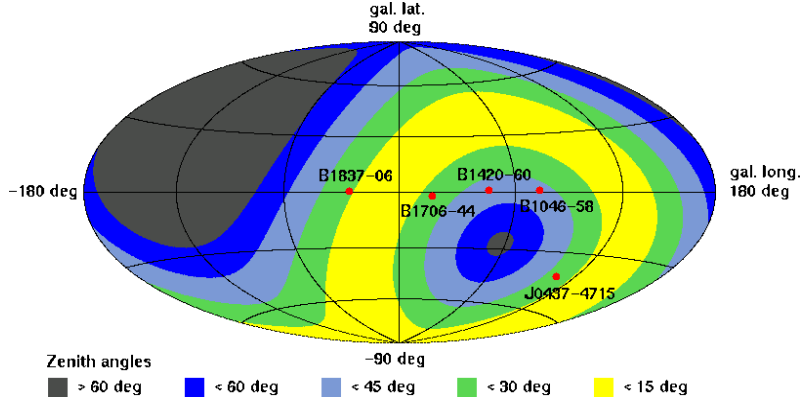,width=10.5cm,clip=} }
\caption{H.E.S.S. visibility. Courtesy of S. Gillessen (MPI-K, Heidelberg).
\label{f1}}
\end{figure*}
\section{Millisecond pulsars}
Most of the radio pulsars detected so far have their rotational periods within the range from 
10~ms to 10~s, whereas a few of them form a separate sample of millisecond 
pulsars with rotational periods in the range from 1~ms to 10~ms. $PSR~1937+214$ is 
the first detected millisecond pulsar, and has a rotational period of 1.5~ms (Backer et al. 1982). 
For such a short period it has a surprisingly small value of $\dot{P}\simeq 10^{-19}$. Using the 
relation given by Eqn.(\ref{tau}), one can estimate the pulsar age as $\tau \geq 10^{7}$~years. 
Being rather old, the millisecond pulsars do not fit the general scheme of pulsar evolution, 
and most probably are re-accelerated by some external mechanism (e.g. by accretion 
from a companion star in a binary system). The millisecond pulsars should have smaller surface 
magnetic field (see Eqn.(\ref{field})), $B_s \sim 10^9~\rm G$, and that makes them particularly 
interesting with respect to the high-energy $\gamma$-ray emission. Given such a low magnetic field, 
the cut-off energy of curvature photons due to the pair production   
(see discussion in section~2) shifts to higher energies, and the maximum radiation 
of the millisecond pulsars should correspond to photons with energy $\sim 10^{11}$~eV (Usov 1983). 
The value of the photon flux above 100~GeV, estimated for $PSR~1937+214$ by Usov (1983), is approximately 
an order of magnitude higher than that of the Crab pulsar, $J_\gamma(\geq 100~\rm GeV)\simeq 2\times 10^{-10}\, \rm ph\, 
cm^{-2}\, s^{-1}$. Recent detailed calculations of the energy spectra of high-energy and very high energy 
$\gamma$-ray emission from millisecond pulsars (Bulik, Rudak, Dyks 2000) in general have confirmed early 
expectations. 

Pulsar $J0437-4715$ ($P=5.75\,$ms, $d=140$~pc) is a prime candidate for observations in the energy 
range above 100~GeV. Its integral $\gamma$-ray flux above 100~GeV is expected to be as high as 
$\simeq 10^{-11}\,\rm ph\, cm^{-2} \, s^{-1}$ (Bulik, Rudak, Dyks 2000), which is above the 
sensitivity limit of H.E.S.S. (Konopelko 2000). Assuming rather conservative $\gamma$-ray flux 
for $J0437-4715$ one needs a 50~hour exposure with H.E.S.S. in order to detect the pulsed 
$\gamma$-rays. Another millisecond pulsar, $PSR~B1821-24$, is a good candidate for pulsed high-energy 
$\gamma$-ray emission seen by EGRET, and it is certainly a valuable target for future observations 
with H.E.S.S. 

\section{Young pulsars}
In addition to seven well established high-energy $\gamma$-ray pulsars EGRET discovered a large number 
of unidentified $\gamma$-ray sources at low Galactic latitudes (see Kanbach 2002). The recent Parkes 
multi-beam radio surveys allowed to resolve a number of young energetic radio pulsars with some of 
those being in rather good positional coincidence with unidentified EGRET $\gamma$-ray sources 
(Torres, Butt, Camilo 2001). Such associations resemble in many ways the established EGRET $\gamma$-ray 
pulsars. For instance pulsar $PSR~B1046-58$ could be a counterpart of high-energy $\gamma$-ray source 
$3EG~J1048-5840$ (Kaspi et al. 2000). It has the ninth highest value of $\dot{E}/d^2$, and its $\gamma$-ray 
light-curve morphology is similar to other known $\gamma$-ray pulsars, namely, two narrow peaks 
separated by $\sim 0.4$ in the light-curve phase. $3EG~J1048-5840$ has a rather flat energy spectrum ($\alpha\simeq 2.0$)
without clear evidence for a cut-off. Two other young radio pulsars $PSR~J1420-6048$ ($P$=68~ms) and 
$PSR~J1837-0604$ ($P$=96~ms) are located $\sim 10'$ away from the position of the EGRET unidentified sources 
$3EG~J1420-6038$ and $3EG~J1837-0606$, respectively (D'Amico et al. 2001). In the energy range from 100~MeV to 10~GeV, 
the spectra of these two sources can be fitted with power laws of photon spectral index $\alpha =$~2.02 and 
1.82, respectively, which are inside the spectral index range of the known $\gamma$-ray pulsars.  
A 20~hour exposure with H.E.S.S. for this source may be sufficient for a clear 
detection of very high energy $\gamma$-ray emission (de~Jager et al. 2001) from both of these pulsars, 
assuming the energy cut-off at 40~GeV. About 30 young energetic radio pulsars are currently known 
as positional associations with unidentified EGRET $\gamma$-ray sources (D'Amico 2001). In 
addition some of those pulsars are within the X-ray nebula, which confirms that they are plerions (see
Roberts, Romani, Kawai 2001), and may well be in addition the sources of DC high-energy $\gamma$-ray 
emission coming from the X-ray nebula, surrounding the pulsar. Such sources could be effectively studied 
at very high energies from the ground with an instrument like H.E.S.S.   

\section{H.E.S.S.}
H.E.S.S. ({\it High Energy Stereoscopic System}) is one of the next generation ground-based instruments 
for very high energy $\gamma$-ray astronomy (Hofmann 2001). In its 1st phase it will consist of 
four 12~m \v{C}erenkov telescopes deployed in the Khomas Highland of Namibia. H.E.S.S. will have an 
energy threshold in the range of 50 - 100~GeV (Konopelko 2000), an angular resolution for individual 
photons of $0.1^\circ$, and an energy resolution better than 20\%. According to the current 
sensitivity estimate, H.E.S.S. will need a 50~hour exposure to detect a $\gamma$-ray point source with 
photon fluxes $J_\gamma \simeq 10^{-11}\, \rm ph\, cm^{-2} \, s^{-1}$ and $\simeq 10^{-13}\, \rm ph\, cm^{-2}\, s^{-1}$ 
above 100~GeV and 1~TeV, respectively. The first observations with two telescopes are being currently 
planned in early 2003. The complete system is due in 2004. The visibility of H.E.S.S. along with a few 
selected pulsars as the potential targets for future observations is shown in Figure~4. 

\section{Summary}
Pulsars are  objects with an extreme physics environment. They are offering physics processes to be  
studied by plasma electrodynamics, nuclear physics, cosmic ray physics, {\it etc}. Understanding 
the pulsar physics could substantially benefit from the enrichment of the presently very small 
sample of high-energy $\gamma$-ray pulsars. Measurements of energy spectra for already 
established, as well as for newly discovered $\gamma$-ray pulsars, will constrain the modeling 
of high-energy emission and in particular the location of the emitting regions with respect 
to the neutron star. Future observations at high-energies, which can be effectively done with the 
ground-based \v{C}erenkov instruments, will stretch the existing models and will help to 
understand the physics of millisecond pulsars. Observations of associations 
of the energetic young radio pulsars at low Galactic latitudes with unidentified EGRET $\gamma$-ray 
sources will allow to prove that pulsars are in general powerful emitters of high-energy $\gamma$-rays. 
Finally, a rich sample of $\gamma$-ray pulsars can be used for better understanding the conversion 
of pulsar rotational energy into $\gamma$-ray emission, which is determined by the size and geometry 
of emitting regions. 

H.E.S.S. is one of the forthcoming ground-based \v{C}erenkov detectors, such as 
CANGAROO, MAGIC, and VERITAS, and the present discussion is entirely relevant for the future observations 
with any of those instruments.         

\begin{acknowledgements}
I would like to thank Werner Becker, Alice Harding, Okkie de~Jager, Gottfried Kanbach, John Kirk, Yury Lyubarsky, 
Roger Romani, Bronislaw Rudak, Vladimir Usov, and Heinz V\"{o}lk, for discussions on pulsar physics.     

\end{acknowledgements}
   

\end{document}